\documentclass[useAMS]{mn2e}
\usepackage{lscape}
\usepackage{rotating}

\def\Msun{\hbox{M$_\odot$}}

\def\cm3{\hbox{cm$^{-3}$}}

\newcommand{\db}{D{$_{\rm break}$}}		
\newcommand{\nmin}{N{$_{\rm min}$}}

\newcommand{\mgmc}{M$_{\rm gmc}$}	
\newcommand{\rgmc}{R$_{\rm gmc}$}	
\newcommand{\hii}{H{\sc ii}~}

\input psfig.sty
\input epsf.sty
\usepackage{graphicx}
\title[Hierarchical Star-Formation in M33]
{Hierarchical Star-Formation in M33:  Fundamental properties of the star-forming regions}
\author[]{N. Bastian$^1$, B. Ercolano$^2$, M. Gieles$^3$, E. Rosolowsky$^2$, R. A. Scheepmaker$^4$,  \newauthor R. Gutermuth$^2$, Yu. Efremov$^5$\\
$^1$ Department of Physics and Astronomy, University College London, Gower Street, London, WC1E 6BT, UK\\
$^2$ Harvard-Smithsonian Center for Astrophysics, 60 Garden Street, Cambridge, MA 02138, USA \\
$^3$ European Southern Observatory, Casilla 19001, Santiago 19, Chile  \\
$^4$ Astronomical Institute, Utrecht University, Princetonplein 5, NL-3584 CC Utrecht, The Netherlands\\
$^5$ Sternberg Astronomical Institute of Moscow State University, Universitetsky Prospect, 13, Moscow, 119899, Russia
}
\date{Accepted. Received; in original form}
\pagerange{\pageref{firstpage}--\pageref{lastpage}}
\pubyear{2005}
\begin{document}
\maketitle
\label{firstpage}
\begin{abstract}

Star-formation within galaxies appears on multiple scales, from spiral structure, to OB associations, to individual star clusters, and often sub-structure within these clusters.  This multitude of scales calls for objective methods to find and classify star-forming regions, regardless of spatial size.  To this end, we present an analysis of star-forming groups in the local group spiral galaxy M33, based on a new implementation of the Minimum Spanning Tree (MST) method.  Unlike previous studies which limited themselves to a single spatial scale, we study star-forming structures from the effective resolution limit ($\sim20$~pc) to kpc scales.  Once the groups are identified, we study their properties, e.g. size and luminosity distributions, and compare them with studies of young star clusters and giant molecular clouds (GMCs).  We find evidence for a continuum of star-forming group sizes, which extends into the star-cluster spatial scale regime.  We do not find a characteristic scale for OB associations, unlike that found in previous studies, and we suggest that the appearance of such a scale was caused by spatial resolution and selection effects.  The luminosity function of the groups is found to be well represented by a power-law with an index, $-2$, the same as has been found for the luminosity and mass functions of young star clusters, as well as the mass function of GMCs.  Additionally, the groups follow a similar mass-radius relation as GMCs.  The size distribution of the groups is best described by a log-normal distribution, the peak of which is controlled by the spatial scale probed and the minimum number of sources used to define a group.   We show that within a hierarchical distribution, if a scale is selected to find structure, the resulting size distribution will have a log-normal distribution.   We find an abrupt drop of the number of groups outside a galactic radius of $\sim4$~kpc (although individual high-mass stars are found beyond this limit), suggesting a change in the structure of the star-forming ISM, possibly reflected in the lack of GMCs beyond this radius.  Finally, we find that the spatial distribution of H{\sc ii} regions, GMCs, and star-forming groups are all highly correlated.

\end{abstract}
\section{Introduction}\label{intro}

Studies of star-forming regions in galaxies have found evidence for a
fractal or hierarchical nature of the distribution of star-forming
sites, similar to that of the ISM (e.g. Elmgreen \& Salzer~1999).  This
complicates the somewhat more simple picture of star-formation that
occurs in either compact star clusters or loose associations.  The
terminology used to describe star-forming groups (young massive
clusters, OB associations, scaled-OB associations or SOBAs) reflects
the idea that multiple distinct entities exist and are fundamentally
different from each other.  However, nearby OB-associations
(e.g. Orion OB1) often contain sub-structures (Blaauw~1964) and multiple distinct clusters along with a general background of star-formation (Elmegreen et al.~2000).
Young clusters in turn are often made up of distinct sub-clusters (see review in Elmegreen~2006a).   Even within the extreme environments of merging galaxies, young star clusters appear as part of larger groups, or complexes, which in turn are often part of even larger structures (Zhang, Fall, \& Whitmore~2001, Bastian et al.~2006).  Thus, evidence is rather pointing to a continuum of star
formation sites.  It is within this context that we ask the question, what are the demographics of star-formation?

The hierarchical  clustering of  young luminous stars was first noted  by Efremov (1984) inside the   30 Dor  super association (by eye) and for the whole LMC  by Feitzinger and Braunsfurth (1984),  using  an objective method.  The issue of whether or not a preferred scale exists for groupings of young stars has been widely discussed since then.  For example,  Efremov (1995),  Bresolin et al. (1998), and Pietrzy{\'n}ski et al.~(2001) have found a preferred size of about  $\sim100$~pc (diameter) for 
OB-associations, whereas Elmegreen \& Efremov (1996, 1998)\footnote{In the latter paper, existing  as e-preprint only, the complete history of the issue is given.}  and Efremov  \&  Elmegreen (1998) concluded that  no such characteristic  scale exists,  the observed one being nothing but an artifact of selecting stars of more or less similar age.  

 It has been shown that the molecular gas within galaxies displays scale-free mass and size distributions, suggestive of a hierarchical nature (e.g. Elmegreen \& Falgarone~1996).  Thus, if star-formation passively traces the gas distribution we would expect young stars to also be organised in a scale-free/hierarchical way, hence most star-formation sites should not be isolated.  Embedded young clusters in the Galaxy appear to show multiple scales of correlation, making it difficult to define a single 'cluster' or  where the 'cluster' ends and the general background of star-formation begins (e.g. Gutermuth et al.~2005).  Again, these star-forming regions are generally not isolated but part of        
larger structures, where the size of these larger structures is defined by      
the observer when they judge which groups appear correlated, i.e. appear to     
have a similar age within some defined tolerance (Efremov \& Elmegreen          
1998).

Humphreys \& Sandage~(1980) identified 143
associations in M33 and found a mean diameter of 200~pc.  However Ivanov~(2005) suggested a characteristic size (diameter) of OB associations in M33 of $60-100$~pc,  suggesting
that resolution effects may play an important role.  Part of the debate on a characteristic size scale is due to a lack of a universal definition of OB-associations and clusters, and where boundaries between these two groups (and larger collections of OB-associations) are drawn (e.g. Elmegreen \& Efremov 1998).  In this work we will use the same terms as laid out in Battinelli, Efremov, \& Magnier~(1996) and references therein.  Groupings of sources found on any scale will be generally termed "groups".  More specifically we will refer to groups with sizes (radii) less than 15~pc as clusters, groups with sizes 
$\sim15-100$~pc as associations, and groups with sizes of a few hundred pc as complexes.   However, as will be shown, these definitions are somewhat arbitrary and the observations suggest a continuum of star-formation sites.

In order to address the question of the demographics of star-formation, we have further developed an objective method to identify and categorise star-forming regions, regardless of scale.  Our approach
is scale independent and therefore should be able to identify
structure within structure (i.e. a hierarchy) if present.  Based on
Minimum Spanning Trees, which have already been extensively used to
study star-forming groups (e.g. Battinelli ~1991, Bresolin et al.~1998, Pietrzy{\'n}ski et al.~2001), we have furthered the general technique,
which avoids the pitfalls of choosing a characteristic size scale for
the search.   We choose the nearby spiral galaxy M33 as our first case
study, presented here.  Its proximity allows individual massive stars to be observed from the ground, and a recent survey (Massey et al.~2006) has provided full coverage of the entire optical disk.  At the distance of M33 (assumed here to be $\sim800$~kpc; Lee et al.~2002, somewhat closer than the 840~kpc found by the HST Key project - Freedman et al.~2001) one arcsecond corresponds to $\sim3.9$~pc.

Finally, Ivanov~(2005)
studied the distribution of OB stars in M33 and found that 70\% of
these stars are associtated with the centres of \hii regions,
confirming the gas rich environments that young stars find themselves in.
In the present work we will also compare the distribution of the
star-forming sites, \hii regions and the giant molecular clouds. 

This work is organised in the following way:  in \S~\ref{sec:data} we introduce the dataset and selection criteria used to select young massive (OB) stars.  In \S~\ref{sec:mst} we introduce the method (fractured Minimum Spanning Trees) to identify and study the distribution of the OB stars.  The size distribution and the idea of a characteristic scale of OB associations are discussed in \S~\ref{sec:size}, while in \S~\ref{sec:results} we present the results and discussion, in particular the relation between the found groups and dense clusters as well as giant molecular clouds.  In \S~\ref{sec:fractal} we present models of a fractal distribution of stars and the resulting size distribution when using the fractured Minimum Spanning Tree method.  Finally, in \S~\ref{sec:conclusions} we summarise our main results.

\section{Observations: M33}\label{sec:data}

\begin{figure}
\hspace{-0.5cm}
\includegraphics[width=9cm]{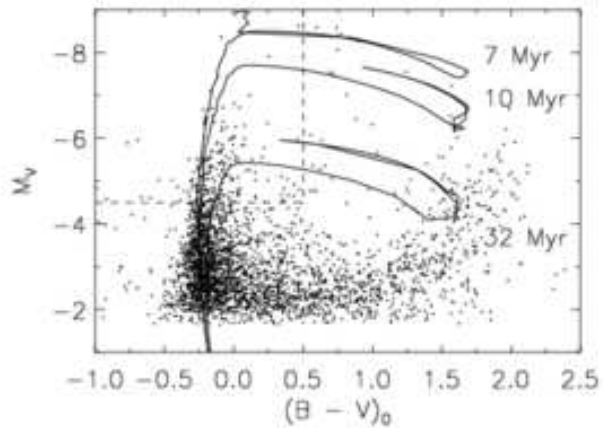}
\caption{The colour-magnitude diagram of a star-forming region south of the nucleus of M33.   The dashed box shows our selection criteria.  Stellar isochrones for three different ages are shown.  }
\label{fig:cmd}
\end{figure}

We extracted our M33 dataset from the UBVI ground-based survey of
local star-forming galaxies, recently published by Massey et
al.~(2006).  We refer the reader to this paper for details on the
observations, data reduction, and photometry\footnote{The data are
  available at: \\ http://www.lowell.edu/users/massey/lgsurvey.html.}.
Due to the proximity of M33 we were able to probe to relatively faint
absolute magnitudes.   We corrected for foreground galactic extinction, assuming 0.227,
0.181, 0.139 and 0.081~mag in the U, B, V, and I bands, respectively
(Schlegel et al. 1998).  In order to focus on young star-forming regions we
applied colour and magnitude cuts of  $(B-V)_0 <$ 0.5 and M$_{\rm V} <
-4.5$~mag.  These cuts were chosen to mimic previous studies (e.g. Bresolin et al.~1998).  An example colour-magnitude diagram of a
star-forming region in M33 is shown in Fig.~\ref{fig:cmd}, where the
dashed box marks the boundaries corresponding to our colour and
magnitude selection criteria.  Additionally, we show three stellar isochrones from the Padova models (Girardi et al.~2002 and references therein) of solar metallicity and ages of 7, 10, and 32 Myr. Our reddening correction does not include a contribution from internal extinction within M33. Massey et al.~(2007) find that an additional A$_V$$\sim$0.22 should be included on average to take the latter into account, however this would be only valid as a statistical correction, given that the correction for any individual source is uncertain. The additional extinction term would effectively raise our faint magnitude limit from M$_{\rm V} = -4.5$ to $\sim-4.3$~mag as well as shift our colour-cut $\sim0.07$ mag to the red, allowing a relatively small number of sources with redder colours to enter our selection box.

\begin{figure}
\hspace{-0.5cm}
\includegraphics[width=9cm]{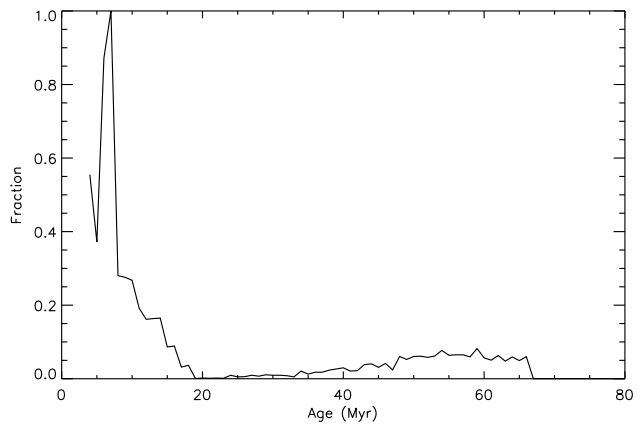}
\caption{The fraction of stars which pass our selection criteria
  assuming a continuous star-formation rate, a Salpter (1955) IMF and
  solar metallicity, normalised to the peak value.  Using our
  selection criteria ($(B-V)_0 < 0.5$ and $M_{V} < -4.5$) 65\% of the
  selected sources are expected to be younger than 20~Myr.  This increases to  73\%, 89\%,  and 98\% if the extinction in each group is $A_{V} = 0.25, 0.5$, and $0.75$ mag respectively.}
\label{fig:fraction-cmd}
\end{figure}

The constant cut in magnitude corresponds to a variable cut in both
initial and present-day mass of the stars, depending on their ages.
For example, the minimum (initial) stellar mass found with the chosen
selection criteria (assuming no extinction) is 18.5, 15.7 and 8.8
\Msun~ for the three ages shown in Fig.~\ref{fig:cmd}.  The fraction
of stars which pass our criteria (assuming a continuous star-formation
rate, a Salpeter~(1955) IMF, and solar metallicity) normalised to the
peak value as a function of age is shown in
Fig.~\ref{fig:fraction-cmd}.   For a continuous star-formation rate
within a given region, 65, 73, 89 and 98\% of sources are expected to
be younger than 20~Myr, assuming extinctions (A$_{V}$) of 0, 0.25, 0.5 and 0.75 mag, respectively. Thus our selection criteria limit us to studying massive star-forming regions younger than 70~Myr old, although preferentially selecting  young, low-extinction regions.  

One important caveat to note is that extinction may move
sources out of our selection box.  Due to the present colour cuts (see
Fig.~\ref{fig:cmd}) only groups with relatively low extinction are
found.  However, an expansion of our selection box (in
colour-magnitude space), automatically implies the inclusion of a
larger number of older ($>$50-100~Myr) sources with low extinction,
adding noise to our results. 

 An additional complication to our selection criteria is posed by the presence of foreground stars. Massey et al.~(2006) estimate that $\sim40$\% of stars with (B--V)$_{\rm 0}$ between 0.3--1.0 and V$_{\rm 0}$=14.6--19.6 are foreground objects.  However, these contaminating stars should not significantly affect our results for two reasons.  First of all, many (perhaps the majority) of foreground stars in the range quoted above fall outside our selection box.  This is shown in Fig.~\ref{fig:cmd} where there does appear to be contamination in the bottom right of our selection box, however most sources fall just to the right of the selected region.  We therefore expect that the majority of the objects in our sample are true high-mass young stars associated with M33.  A second reason is that foreground stars should be spread randomly over the field. The algorithm we use to selects groups of stars (described in the next section) requires any group to be composed by a minimum number of sources, N$_{\rm min}$.  In this paper we adopt N$_{\rm min} = 5$, drastically reducing the probability of chance alignments of field stars to dominate a group's population.  We can therefore conclude that, in the worst case scenario, the addition of a random distribution of "non-legitimate" sources to our dataset would only have the effect of raising the noise level of our results, without causing a bias in a particular direction.

\section{Fractured Minimum Spanning Trees (fMST)}
\label{sec:mst}

In order to objectively study the spatial distribution of young stars in M33
we constructed a Minimum Spanning Tree (MST) of all the sources that
pass our selection criteria in the galaxy.  The MST method connects
all points to their nearest neighbour such that the total length of
all connecting segments is minimised without forming any closed loops, and all sources are connected to the same tree.
The resulting MST can then be {\it fractured} by applying a breaking
criterion \db, whereby all connecting segments longer than \db~ are
removed, leaving us with a sample of isolated groups.  The question of
how to choose \db~ has been considered previously, with the aim of
setting an unambiguous and reproducible length scale.   The most
commonly applied \db~ is that described by Battinelli ~(1991), in which
a histogram is constructed of the number of associations (i.e. groups) found as a function of \db~(for a given minimum number of sources which defines a group).  The \db~ chosen for the analysis is set to
the distance in which the histogram peaks.  This technique has the
advantage of providing a uniquely determined \db, but is severely
limited by it being resolution-dependent, as will be described below.

In order to prevent an artificial length scale from entering our data,
we have fractured our MST of M33 using multiple \db, from 19~pc to
252~pc in 13 equal steps.  The lower value
was chosen according to the resolution limit of the data (i.e. smallest distances where multiple sources could be resolved).   This
allows us to see all levels of the hierarchy and compare our results
directly with other studies.  An additional parameter to be considered is the minimum number of sources which is required to identify a region, \nmin.  In this work, we will use \nmin~=5 unless otherwise stated.  However, when appropriate, we will describe the effect of different values of \nmin~ on our conclusions.

Once the groups are isolated (i.e. by applying the adopted \db) their
centres are defined as the mean of the position vectors marking the
locations of all source belonging to a given group.  In most groups
the centres calculated by our algorithm fall on or around a
potentially eye-estimated centre. A small fraction of the derived
groups, however, show extremely irregular structures, such as (e.g.)
long, thin tails, causing the calculated centre to fall outside the
main body of the group.  The radii of the groups are defined to be
equal to half that of the distance between the two furthest sources in
the group, regardless of whether the connecting line passes through the adopted centre.

\begin{figure}
\hspace{-0.5cm}
\includegraphics[width=8cm]{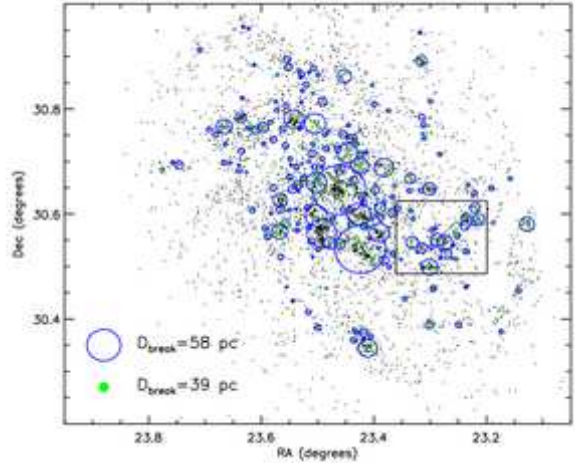}
\caption{The positions of all the sources that passed our criteria are shown as black points.  Groups found with the \db= 58~pc criterion are shown, along with a circle representing their radii.  Groups found with the \db =39~pc criterion are shown as green dots.   The boxed region is shown in more detail in Fig.~\ref{fig:image2}.} 
\label{fig:image1}
\end{figure}

\begin{figure}
\hspace{-0.5cm}
\includegraphics[width=8cm]{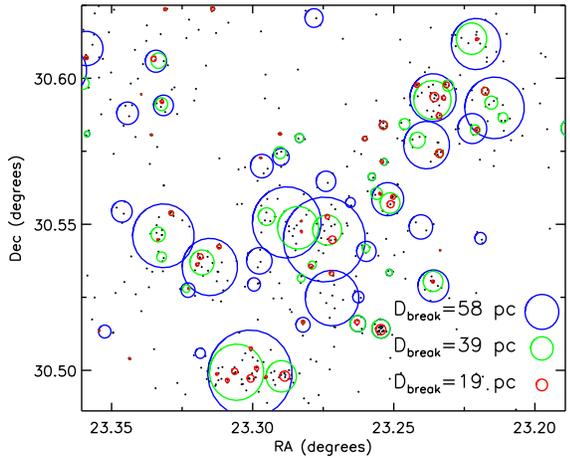}
\caption{A blowup of the boxed region shown in Fig.~\ref{fig:image1}.  Here \nmin = 3 to highlight the number of small groups.} 
\label{fig:image2}
\end{figure}

Figure~\ref{fig:image1} shows all of the groups found with five or
more sources for \db = 58~pc (circles, blue) and \db = 39~pc (filled
green dots).  The size of the circle represents the derived radius of
each region.  We note however that this is just a representation of
the actual derived groups, as few groups are circular.  The boxed region is shown in more detail
in Fig.~\ref{fig:image2}, where we show all the groups found for
three \db~ values.  The sources which appear to only belong to the lower levels of the hierarchy (i.e. \db = 19 or 39~pc) actually were also found by the higher levels, however only the lowest level is shown. 
By the nature of the region finding routine, any group found by the smallest search size is automatically found by the larger search sizes, with equal or larger radius.  The derived groups can overlap in Figs.~\ref{fig:image1}~\&~\ref{fig:image2} only because of their representations as circles.  

A total sample was then constructed by combining all thirteen \db~
and removing groups which were found by multiple \db.  In particular,
if a group had exactly the same position and radius between different
\db values, it was only counted once.  Smaller groups which were found inside
larger ones were kept in order to preserve the signature of a
hierarchy.  When results from the individual \db~ are
discussed, all groups are included, regardless of whether they were
identified by multiple \db values (i.e. each \db~ is treated independently of all others).

\section{Size distribution}
\label{sec:size}

Once the regions have been found and catalogued, we can begin studying
their general properties in order to determine their demographics and compare them to gas/dust clouds in the ISM from which they presumably formed.

\begin{figure}
\includegraphics[width=8cm]{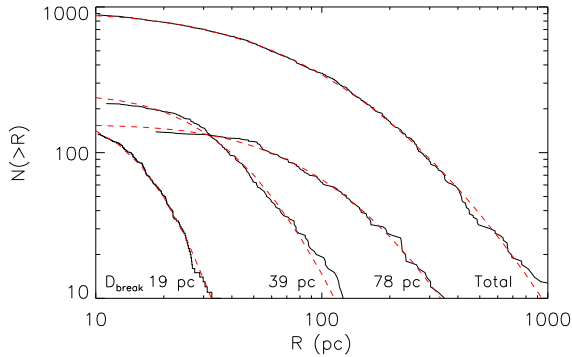}
\caption{The size distribution derived from the M33 sample.  Only
  groups with five or more members (\nmin=5) are included in the
  figure.  The results for three different \db~ are shown, along with
  the total sample (i.e. the combination of all \db~ with the doubles
  removed).   The dashed (red) lines represent log-normal fits to the
  data, which do an excellent job in reproducing the observations.  In
  this data representation, a power-law distribution would be a straight line, which clearly does not fit the data.}
\label{fig:rad-dist}
\end{figure}

Figure~\ref{fig:rad-dist} shows the cumulative radius distribution of the star-forming groups found using different breaking distances, \db, as well as the total sample (i.e. where each region has only been counted once).   All results shown in Fig.~\ref{fig:rad-dist} only use groups in which five or more sources were found, i.e \nmin=5.  The effect of \nmin~ on the results will be discussed in \S~\ref{sec:preferred}.  In order to quantify the results we fit a log-normal function of the form:
\begin{equation}
f(R) = \frac{p_0}{R p_2 \sqrt{2\pi}} \exp\left[-\left(\frac{(\ln R - p_1)^2}{2p_2^2}\right)\right] 
\end{equation}                                                       
 which corresponds to a cumulative distribution of:
 \begin{equation}
 N(>R) =\frac{p_0}{2} \left[1-\mbox{erf} \left(\frac{\ln R-p_1}{p_2\sqrt{2}}\right)\right],                                                  \end{equation}
 where $R$ is the radius of each group, and $p_0$, $p_1$, and $p_2$ are the number in the distribution, the natural log of the mean value in the sample, and the dispersion in e-foldings, respectively.
  
  The dashed lines in the plot represent log-normal fits to the data.  In this representation, a power-law distribution of the form $NdR \propto R^{-\eta}dR$ would be a straight line in the plot, which clearly does not accurately represent the data.   Power-laws are often used as parameterised fits to data, in particular size distributions, however it is difficult to differentiate between a power-law and a log-normal function unless a sufficient range (generally greater/approximately than two dex) is covered by the variable which is being fit.

We will discuss the significance of this result is Sections~\ref{sec:star-clusters} \&~\ref{sec:gmcs}. 

\subsection{A preferred size?}
\label{sec:preferred}

\subsubsection{The role of the breaking distance}

Previous studies, e.g. Bresolin et al.~(1998), have suggested a
characteristic distance scale in the size distribution of OB
associations.  Bresolin et al.~(1998) have estimated this scale to be $\sim40$~pc in
{\it radius} based on studies of spiral galaxies between 6 and 14~Mpc
away, using HST-WFPC2 data, and a similar technique as the one applied
here.  The technique adopted for the above study, was the prescription given in Battinelli ~(1991), namely the construction of a minimum spanning tree broken using multiple values of \db.  However, their approach to the problem was fundamentally different from ours, in that the multitple \db~values (or in their terms "search radii") were applied in order to determine a {\it preferred} \db, corresponding to the value which returned the largest number of groups.  Only the preferred \db~ found in this way was then used in their further analysis.

The use of multiple \db~ values puts us in a favourable position with respect to
previous studies that used a single \db.  The data in Tables~1~\&~2 of Bresolin et
al.~(1998) clearly shows that the mean radius of the groups of which they find within each galaxy and their \db~("search radius"), change as a function of the distance to the galaxy, suggesting that resolution effects may be at play.

\begin{figure}
\includegraphics[width=8cm]{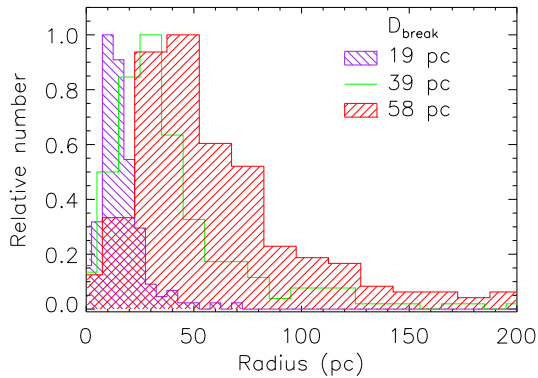}
\caption{The size distribution derived from the M33 sample for three different \db.  Only groups with five or more members (\nmin=5) are included in the figure.  Note that the turn-over in the distributions shifts to smaller sizes for smaller \db.}
\label{fig:rad-his}
\end{figure}

In order to compare our results to previous studies, we show the size
distribution of regions in M33 for three different \db~
(Fig.~\ref{fig:rad-his}).  A turnover in the size distribution is
clearly seen in our figure; the radius at which this turnover occurs
decreases for smaller \db values. It is interesting to note that a
39~pc breaking bistance (which mimics that used in the Bresolin et
al. (1998) sample), results into a mean radius of $\sim46$~pc,
consistent with average radius of 40~pc, as
measured in their study. Additionally, we note that the mean (i.e. $e^{p_1}$ in the functional form fit to the cumulative radius distribution) increases in an almost one-to-one way with increasing \db.

The lowest value of \db~ used ($19$~pc) in the current study is dictated by the resolution
limit of the catalogue ($\sim5$") in order to detect at least five bright sources.  This implies that, naturally, our
fractured MST (fMST) 
algorithm is unable to pick up the large number of young stars
contained in dense clusters below our resolution limit, as these will
all be blended and appear as single sources. Presumably, if higher resolution data is used the size distribution of the star-forming groups would continue to much smaller sizes, running seamlessly into the star cluster distribution.  It should be noted, however, that the star cluster distribution does appear to have a preferred size, with a turn-over in the radius distribution at $2-4$~pc (e.g.~Bastian et al. 2005a, Jord{\'a}n et al.~2005, Scheepmaker et al.~2007).  We will return to this point and its significance in \S~\ref{sec:star-clusters}.


\subsubsection{The role of the minimum number sources used}

As discussed above, previous studies have concluded that
there exists a characteristic scale of star-formation, namely that of
the OB associations, typically with radii of 40--60~pc.  However, we
showed above that a combination of a chosen \db~ and resolution effects, can mimic a characteristic scale length.  Here we investigate the role of the adopted minimum number of sources required to define a region, \nmin.

As described in \S~\ref{sec:mst} the use of multiple \db~ should
preserve the signature of the hierarchical nature of star-formation, if present.   Thus, the analysis adopted here should have no preferential size associated with it, except in regard to the relation between the number of sources within a region and its size (which will be discussed in \S~\ref{sec:mass-radius}).

\begin{figure}
\includegraphics[width=8cm]{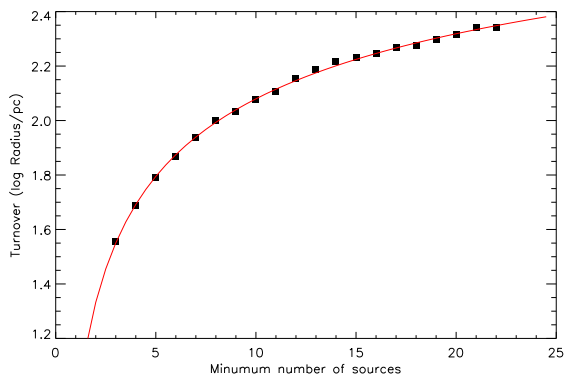}
\caption{The turn-over in the radius distribution as a fuction of the minimum number of stars used to define a group.   All groups have been combined (i.e. the total sample has been used, see text for details).  The plot shows the distance at which the gaussian fit of the size distribution peaks (e.g. as shown in Fig.~\ref{fig:rad-dist}), plotted against \nmin.  The squares represent the results for M33.  The solid (red)  line is a  geometric fit to the data.  Clearly, as the number of sources drops so does the turn-over.}
\label{fig:turnover}
\end{figure}

Following on the results of the previous section, we constructed
histograms of the radii (binned in logarithmic steps) for different
selection criteria on \nmin, running from 3 to 25 sources.  We then
fit a Gaussian to the distribution and the size at which the Gaussian peaks
vs. \nmin~is shown in Fig.~\ref{fig:turnover}.  Clearly, as \nmin~
decreases so does the turnover radius.  The solid (red) line in the
figure is a geometric fit to the data for M33.  Extrapolating to
smaller values of \nmin~we see that the turnover falls below ~10~pc,
the regime of compact clusters.  The mean and median of the
distributions both show the same behavior as shown in
Fig.~\ref{fig:turnover}, suggesting that the values of mean and median
diameters found previously for OB associations were controlled by the selection criteria and resolution of the data.

Thus we are left to conclude that no preferential scale exists
for OB associations, or star-forming groups.  This is not necessarily at odds with observations of H{\sc ii} groups in galaxies, such as M51  which shows a turnover in their size distribution at 10--15~pc in radius (Scoville et al.~2001).   Due to the size-luminosity relationship found by Scoville et al.~(2001) for the H{\sc ii} groups, smaller groups are less luminous, hence closer to the detection limit.  If these groups have the same average extinction as found for the larger groups (A$_{V}$=2.5) they will be shifted below the detection limit, and hence not appear in the extinction corrected sample.  In fact, the extinction corrected luminosity distribution of H{\sc ii} groups in M51 has a steep decline at L$_{\rm H\alpha} = 10^{37}$~ergs/s, which corresponds to a diameter of $\sim25$~pc (Scoville et al.~2001), exactly at the point of the turnover, suggesting that incompleteness is playing a crucial role.

In \S~\ref{sec:fractal} we will further discuss the implication of the log-normal distribution and how it naturally occurs when selecting a specific size scale to study within a hierarchical distribution.


\section{Results}
\label{sec:results}

\subsection{Luminosity function}
\label{sec:lum}

Studies of young star clusters, from the relatively quiescent solar
neighbourhood to extreme starbursts have shown that they form with a
power-law mass function (MF) of the form $NdM \propto M^{-\alpha}dM$, with $\alpha \sim 2.0$ (e.g. Bik et al.~2003; de Grijs et al.~2003).  However, what is usually measured as a proxy for the mass function is the luminosity function, LF.  In the presence of a (almost) single age population the LF is equivalent to a MF.  Studies of full cluster populations often find a steeper LF index, $\alpha=2.0-2.4$ which may be due to an age spread and an age-dependent extinction (Larsen~2002) if the underlying mass function is truncated (Gieles et al.~2006).

\begin{figure}
\includegraphics[width=8cm]{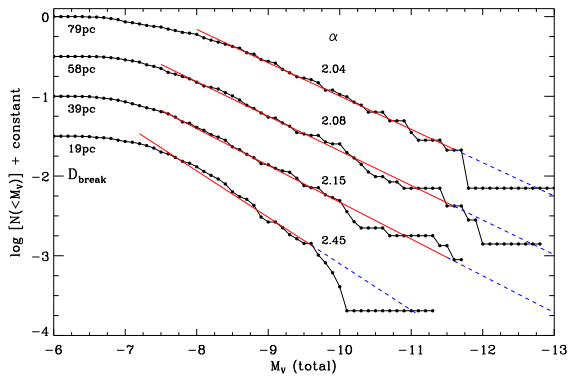}
\caption{The cumulative luminosity distribution of groups in M33 for four different \db.  M$_{\rm V}$~(total) represents the magnitude of all the sources identified within the group which pass our selection criteria.  In this representation, assuming a functional form of $N(L)dL \propto L^{-\alpha}dL$, the index can be found by $\alpha = 2.5 *$slope$ + 1.$} 
\label{fig:cum-lum}
\end{figure}

Looking at the luminosity function of stellar concentrations in NGC~628, Elmegreen et al.~(2006) find a power-law distribution with index, $\alpha \sim2$.  In order to compare our results with previous works, we construct cumulative luminosity distributions of star-forming regions in M33 for different \db.  We then fit the slope (in a log-log plot) of the distribution, which relates to $\alpha$ as 2.5*slope + 1.  An example of the routine is shown in Fig.~\ref{fig:cum-lum} were we show the luminosity distribution for four \db.  The solid lines (red) show the best fit to the data over regions where the fit was carried out.  The dashed lines (blue) indicate an extension of the best fit to larger luminosities. For the four values shown, we find an average index of  $\alpha = 2.15\pm0.13$, in good agreement with previous results of stellar concentrations (i.e. groups) and star clusters.  The errors were found by deriving $\alpha$ for the four \db~ shown for different values of \nmin~ between 5 and 10.  We note that the apparent increase in $\alpha$ for smaller \db~ is largely due to the \nmin~ chosen, since smaller regions tend to have fewer sources, and hence lower total magnitudes (see \S~\ref{sec:mass-radius}).

Figure~\ref{fig:cum-lum} appears to have a shallower slope (i.e a turn-over) below $M_V \sim-7.4$ than what would be expected from a continuous power-law for all cases shown.  This is mostly likely due to incompleteness effects.  Since we require a minimum of five sources to define a group (i.e. \nmin=5) and we have applied a magnitude limit of $M_V < -4.5$ in our sample selection, all groups which pass our selection criteria must have $M_V < -6.25$.  However, most groups will have at least one source significantly brighter than our selection criteria, pushing this lower limit to brighter magnitudes.  Hence, it is not possible to determine an absolute detection limit, however the occurrence of the turnover at similar magnitudes for all \db~ considered, suggests that it is a completeness limit effect.

\begin{figure}
\includegraphics[width=8cm]{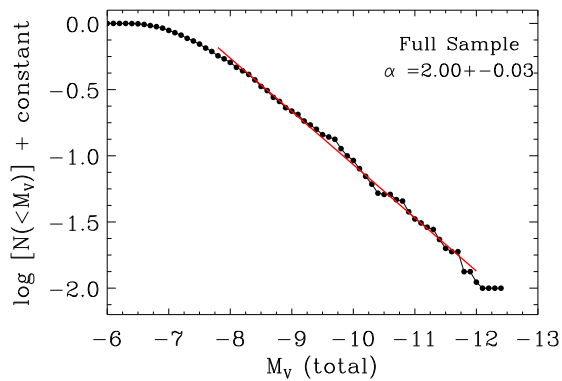}
\caption{The same as Fig.~\ref{fig:cum-lum} but now for the total sample.  The error in the slope was found by varying \nmin~between 3 and 20.  } 
\label{fig:cum-lum-pure}
\end{figure}

We have also looked at the luminosity distribution of all regions
found by the fMST algorithm (eliminating double detections, see
\S~\ref{sec:mst}).  Figure~\ref{fig:cum-lum-pure} shows the results
for \nmin=5, although we note that the slope is independent of this
variable.  Assuming that the luminosity function is an appropriate
tracer for the mass function, we see that the region distribution is
very similar to that observed for young dense star clusters as well as
giant molecular clouds (GMCs).

\subsection{Mass-radius relation}
\label{sec:mass-radius}

If star-forming groupings do passively trace gas structure, then we would expect them to share the same fundamental properties as GMCs, except those which are affected by dynamical evolution or the change from gas to stars.  One such property that is expected to be retained is the observed mass-radius relation of GMCs, \mgmc $\propto$ \rgmc$^2$~(Solomon et al.~1987).  A similar relation was found for eleven large star-forming regions in M51 (Bastian et al.~2005b).  Using the much larger sample found in the present work, we can search for such a relation.  

\begin{figure}
\hspace{-0.5cm}
\includegraphics[width=9cm]{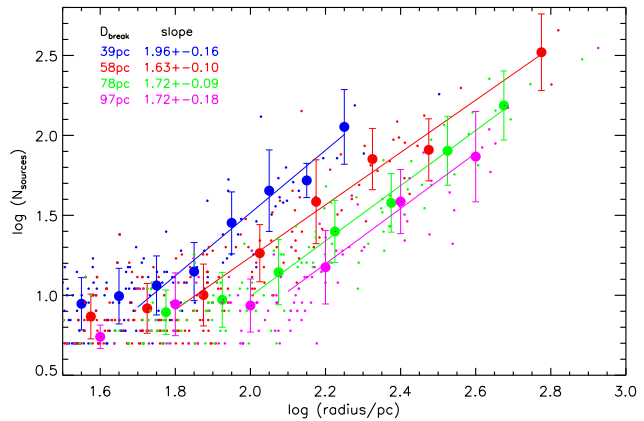}
\caption{The radius of each group vs. the number of sources within the group.  The different coloured points represent the results for different breaking radii.  The large symbols and error bars represent the average number in that radius bin and the standard deviation.  The lines are the best fits through the data, with the slopes for each one given in the panel.} 
\label{fig:rad-n}
\end{figure}

To first order, the number of sources in a group is directly proportional to its mass (assuming all groups are of a similar age), as long as the stellar IMF is fully sampled (i.e. sampling effects are not significant).  In Fig.~\ref{fig:rad-n} we show the measured  radius (logarithm) vs.  number of sources (logarithm) within each group for four \db.  As in the previous figures, the solid lines represent least-square fits to the data where fits where carried out over the interval of the lines. While there is significant scatter in each sample, a similar trend as described in Bastian et al.~(2005b) is found, i.e. if we approximate the relation to a power-law of the form N$\propto R^{\zeta}$, then $\zeta = 1.63 - 1.96$, whereas Bastian et al~(2005b) found $\zeta = \sim2$.

This power-law is expected from a fractal distribution of young stars
(e.g. Elmegreen \& Falgarone~1996).  However, it is worth noting at
this point that a simple random distribution of sources on a two
dimensional plane can also reproduce the same index (i.e. the number of sources increases directly with the surface area).

This is somewhat shallower than found for young stellar regions in the LMC by Gouliermis et al.~(2003), who used a different technique.  Fitting a power-law to their data gives $\zeta > 2.5$, possibly indicating a galactic dependence on the star-forming region properties.



\subsection{Relation to star clusters}
\label{sec:star-clusters}

Based on HST-WFPC2 data, Bastian et al.~(2005a) showed that the distribution of cluster radii in M51 can be approximated by a power-law of the form $NdR \propto R^{-\eta}dR$ where $\eta = 2.2$.   However, using higher resolution HST-ACS images and a much larger sample,  Scheepmaker et al.~(2007) find that a single power-law does not accurately represent the data. A log-normal distribution, like the one presented here, is a much better fit to the data.   The cluster population in M51 shows a distinct peak in the size distribution (when using bins of equal width on a              
logarithmic scale) at $\sim3$~pc.  A similar peak has also been found for young stellar clusters in M101 (Barmby et al.~2006).

As shown in \S~\ref{sec:size} the size distributions of star-forming groups in M33 are also well fit by a log-normal distribution, where the turn-over, or mean size, is determined by the \db~ and \nmin~adopted.  Elmegreen (2006a,b) has proposed that dense star clusters simply represent the dense inner part of a continuous hierarchy of star-formation within galaxies.  Thus, we conclude that the size distribution of the groups in M33 extends into, and in fact becomes, the size distribution of clusters, i.e. that clusters are simply compact groups.  Further support for this conclusion was given in \S~\ref{sec:lum}, where  it was found that the groups in M33 also share the same luminosity distribution as compact star clusters, suggesting a similar, if not identical nature.

One observed difference between the compact clusters and the more extended regions discussed here, is that clusters are known to lack a strong mass-radius relation (e.g. Larsen 2004) whereas groups have a similar mass-radius relation as GMCs.  This difference may be explained due to dynamical evolution of clusters, which due to their smaller sizes, and hence smaller crossing-times, will evolve more quickly (e.g. Elmegreen~2006a,b).  This early evolution may erase any natal mass-radius relation.



\subsection{Comparison of the star-forming regions with the H{\sc ii} region distribution}

\begin{figure}
\hspace{-0.5cm}
\includegraphics[width=8cm]{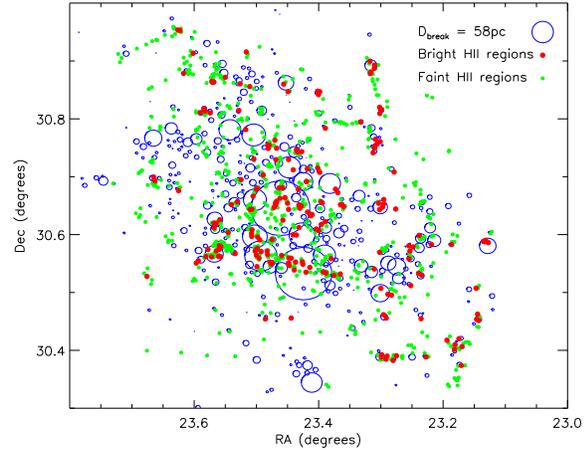}
\caption{The spatial distribution of the star-forming regions (blue circles, \db=58~pc, \nmin=3) as well as bright (large red filled circles) and faint (small green filled cirlces) \hii regions.} 
\label{fig:hii-dist}
\end{figure}

Collections of massive stars are able to ionize the gas surrounding them, thereby creating emission line regions, i.e. \hii regions.  Therefore, a test of the method used here is to compare the spatial distribution of the star-forming regions found with that of the \hii regions.  For this comparison we have compiled a list  \hii of regions from the catalogues of Boulesteix et al.~(1974), Court\`es et al.~(1987) \& Hodge et al.~(1999).  Figure~\ref{fig:hii-dist} shows the positions of the \hii regions (broken into two subsets, faint and bright regions which correspond to 25-300\% and $>300\%$ of the mean \hii region flux respectively) as well as the young stellar groups (\db$=58$~pc, \nmin=3).   The correlation between the bright \hii regions and the stellar groups is excellent, whereas many of the fainter \hii regions are not coincident with any found stellar groups.  

The cause of this is likely that the bright \hii regions are powered
by multiple massive stars, which if they occur in relatively
low-extinction regions, will be found by the fMST algorithm.  The
fainter regions are most likely ionised by only one or two massive
stars (hence failing our selection criterion which requires $N_{\rm
  min}~\ge$~3 for identification of a group), or numerous low-mass stars
which do not pass our colour-magnitude selection criteria.

\begin{figure}
\hspace{-0.5cm}
\includegraphics[width=9cm]{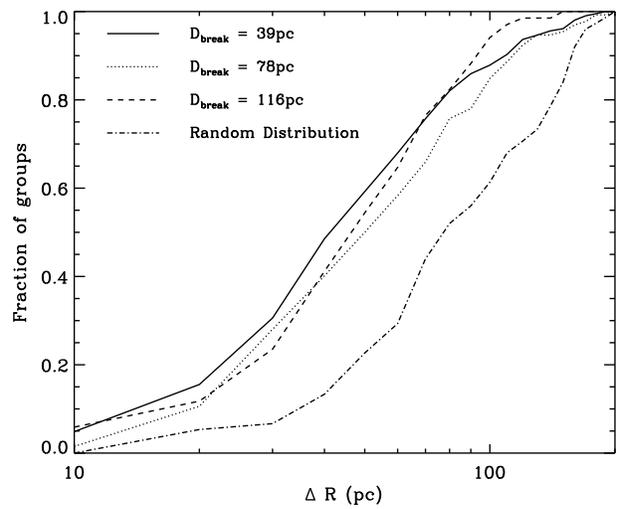}
\caption{The fraction of the groups (for various \db) that have at least one H{\sc ii} region within $\Delta R$ of the group centre.  For comparison we also show the expected relation for a randomly distributed H{\sc ii} region sample (same number of H{\sc ii} regions as the sample used, distributed over an equal area).  The groups and H{\sc ii} regions are significantly correlated over scales of at least 100~pc.} 
\label{fig:hii-comparison}
\end{figure}

In order to quantify the correlation between the groups and H{\sc ii} regions, we follow the same technique as Engargiola et al.~(2003) who compared the distribution of GMCs to H{\sc ii} regions in M33.  Fig.~\ref{fig:hii-comparison} shows the fraction of groups (for various \db) which have at least one H{\sc ii} region within a given distance, $\Delta R$, of the group centre.  The results for a randomly distributed population of H{\sc ii} regions is also shown (dash-dotted line) for comparison.  We find that the two populations (groups and H{\sc ii} regions) are strongly correlated to distances greater than 100~pc.  This is very similar to that found for GMCs in M33 (Engargiola et al.~2003), which will be discussed further below.  It is found that $\sim20$\% of the groups (for the \db$=39$~pc case) have an associated H{\sc ii} region within 20~pc of the group centre.

These results confirm the findings of Ivanov~(2005) who found a good, although not one-to-one, correlation between H{\sc ii} regions and OB stars in M33.  Such a correlation is of course not surprising, since H{\sc ii} regions are powered by ionising flux from massive stars.

\subsection{Comparison with the GMC distribution}
\label{sec:gmcs}

Thus far we have seen that the star-forming regions share some
characteristics with the GMCs.  M33 is an interesting case in terms
of its gas content, as it appears lacking in GMCs ($>
10^{5}M_{\odot}$) at galactic radii larger than $\sim4$~kpc
(Rosolowsky et al.~2007).  Beyond this radius the gas appears to be
much more diffuse without large concentrations.  We can therefore ask
the question: are dense, centrally concentrated, large clouds necessary
to form the star-forming regions found in this study, or are smaller or looser gas clouds adequate?

\begin{figure}
\hspace{-0.5cm}
\includegraphics[width=9cm]{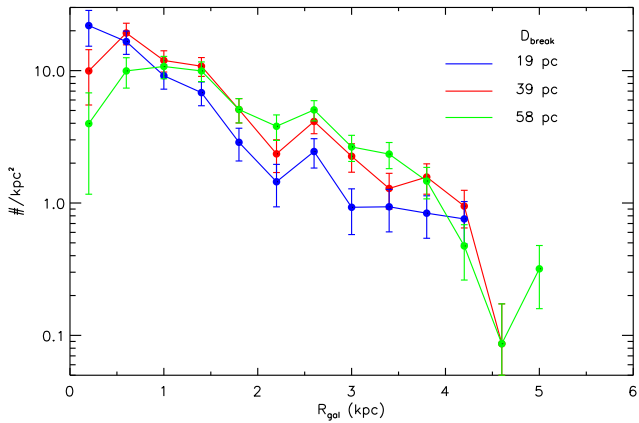}
\caption{The radial distribution of the surface density of regions found for three different \db.  Note the strong drop in surface density at a galactic radius of $\sim4$~kpc, exactly where the GMC distribution is truncated.} 
\label{fig:radial-dist}
\end{figure}

In Fig.~\ref{fig:radial-dist} we show the radial surface density of
regions found for different \db.  The distribution appears centrally
concentrated, similar to the GMC and diffuse molecular gas in the
galaxy (Rosolowsky et al.~2007).  However, all three distributions
(the three \db ~shown) show an abrupt truncation at $\sim4$~kpc,
similar to that shown by the GMCs.  From this we conclude that the
star-forming regions (at least the star-forming regions capable of
producing high-mass stars) trace the distribution and properties of
the dense/high mass molecular gas in M33.   We note that molecular gas             
and H{\sc ii} regions are both found in the outer galaxy, but the absence             
of both large groups of high mass stars and GMCs suggests a sharp               
change in the {\it structure} of the star-forming ISM not revealed in           
radial profiles of the molecular gas or star-formation tracers.

In order to quantify how spatially correlated the star-forming groups and the GMCs are we performed the same test as was carried out for the H{\sc ii} regions, namely by looking at the fraction of groups found with a GMC within a given distance $\Delta R$.  The resulting distribution is very similar to that found for the H{\sc ii} regions, indicating that GMCs and the cataloged groups are correlated on scales of at least 100~pc.  The lower number of GMCs than H{\sc ii} regions, however, does give larger deviations on the random position tests, hence slightly lowering the significance of the detected correlation.

\section{Fractal Structure}
\label{sec:fractal}


If the star-forming regions are distributed in a fractal nature
throughout the galaxy, then the dimension of the fractal can be
retrieved by looking at the number of regions found as a
function of the breaking distance used (e.g. Elmegreen \& Salzer~1999, Elmegreen \& Elmegreen~2001,
Elmegreen et al. 2006)\footnote{Note that this is
  different than what is shown in Fig.~\ref{fig:rad-dist} where the
  size distributions of the regions for a given \db~ is shown.}.  The
results for M33 are shown in Fig.~\ref{fig:cum-vs-db} for the total sample, for three \nmin.  The vertical dashed lines are the critical lengths (defined to be the inverse square root of the mean surface density of sources), where due to the exponential disk of M33, we show the mean density at half the maximum value (left line) and one-fourth the maximum value (right line).  Note that the distribution can be well represented by a power-law above the critical length.  An example of a power-law fit to the \nmin=10 distribution is shown as a solid line (shifted vertically for clarity).

The distributions appear similar to that found by Elmegreen et al.~(2001) for a sample of eight galaxies and Elmegreen et
al.~(2006) for NGC~628, although their technique differs from ours.  Their method was based on smoothing the original image with a gaussian kernel and counting the number of regions found.  This was carried out for seven gaussian kernel sizes from 1 to 64 pixels.
The
similarity between their results for NGC~628 and those presented here
for M33 suggests that their gaussian smoothing length plays a similar role as our \db, locating star-forming structures at a given length scale.  Elmegreen et al.~(2006) presented a series of models of fractal distributions and showed that an intrinsic fractal distribution will have a power-law shape when viewed as in Fig.~\ref{fig:cum-vs-db}.


While such a comparison is not conclusive, it does suggest that the star-forming groups in M33 are not distributed randomly, but instead have structure on a large range of scales, similar to that expected for a fractal distribution.

\subsection{Models and the selection of a size scale}

\begin{figure}
\hspace{-0.5cm}
\includegraphics[width=8cm]{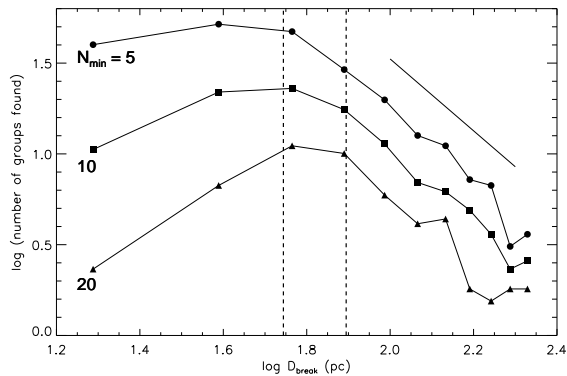}
\caption{The number of groups found (eliminating double counts, i.e. for the total sample) for each \db, and various values of \nmin.  The two vertical dashed lines represent the critical length for two different values of the mean surface density of sources used to construct the MST, namely the value at half (left line) and one fourth (right line) the peak central surface density (see text for details).  The solid line represents a power-law fit to the data for the \nmin=10 case, for \db~values lager than log~\db(pc) = 1.8, shifted vertically for clarity.} 
\label{fig:cum-vs-db}
\end{figure}

It remains to be seen how a fractal or hierarchical distribution of star-forming regions can result in a log-normal size distribution.  The first point to consider is that there are two ways to define a size distribution.  The first, adopted in Fig.~\ref{fig:rad-dist}, is the distribution of group sizes which were found with a given \db~(or the combination of these, as in the total sample).  An equally useful way is to study the number of groups found for a given scale, or \db.  The latter is shown in Figure~\ref{fig:cum-vs-db}.  However, observationally, these two methods result in fundamentally different distributions, the former results in a log-normal distribution while the latter results in a power-law distribution.

\begin{figure}
\hspace{-0.5cm}
\includegraphics[width=8cm]{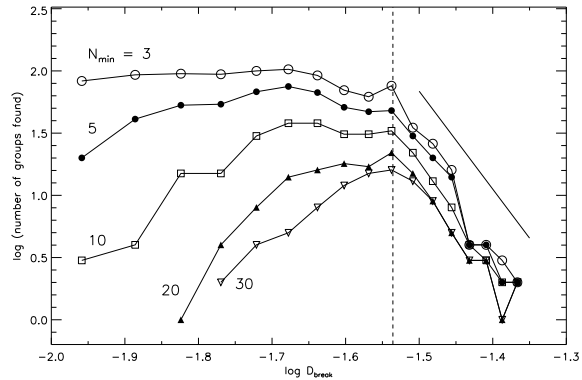}
\caption{The number of groups found (eliminating doubles) in the fractal simulations for different \db.  The different curves represent different values of \nmin.  The dashed vertical line is the critical length of the simulation (see text), while the solid line shows a power-law fit to the \nmin$=10$~curve above the critical length, shifted vertically for clarity.  It is clear that the curves are well represented by a power-laws above the critical length, whereas below this length their form is dominated by the value of \nmin.} 
\label{fig:fractal-sim2}
\end{figure}

In order to understand this phenomenon we have constructed a fractal distribution of sources (stars), using the prescription given in Goodwin \& Cartwright~(2004).  We placed 3000 sources (stars) inside a spherical area with a fractal distribution with dimension D=2.3.  The sources were then projected onto a plane, and the fMST algorithm was run over the model set.  The number of groups found for each \db~ is shown in Fig.~\ref{fig:fractal-sim2}, for five values of \nmin.  The dashed vertical line shows the critical length scale of the simulations, $(N_{\rm total}/Area)^{-0.5}$.  Above this value the distribution is well approximated by a power-law, an example fit (shifted in the y-axis for clarity) is shown as the straight solid line.  Below the critical length, the distributions are controlled by \nmin.  This simple model confirms the prediction of a fractal distribution given in Elmegreen \& Elmegreen~(2001) and Elmegreen et al.~(2006), that such a distribution will have a power-law behavior.

\begin{figure}
\hspace{-0.5cm}
\includegraphics[width=9cm]{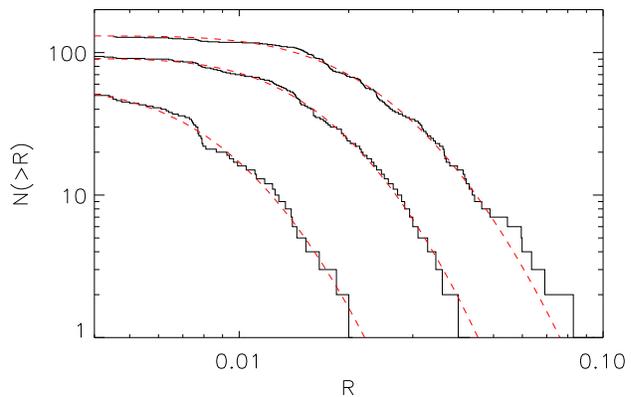}
\caption{The resulting size distribution of groups found in the fractal simulation for three values of \db.  The size distribution of groups selected for each \db~ is well described by a log-normal distribution.} 
\label{fig:fractal-db}
\end{figure}

Additionally, in Fig.~\ref{fig:fractal-db} we show the size distribution of groups found in our simulations for three different \db.  It is clear that these groups follow a log-normal size distribution, just as was found for groups in the M33 sample (e.g. Fig.~\ref{fig:rad-dist}).  Thus we conclude that a hierarchical distribution of sources (i.e. young massive stars in the observations) will give a log-normal size distribution if a particular scale is chosen to study.

This may explain the log-normal size distribution (number of clusters of a given size vs. size) observed for compact clusters (e.g. Jordan et al.~2005, Scheepmaker et al.~2007).  As discussed above, the clusters appear to be the continuation of the hierarchy of the star-forming regions studied here.  Thus, if we choose to study the size distribution of groups in the range of say $1-10$~pc (i.e. clusters) then we would naturally obtain a log-normal size distribution.  Clusters are, however, different than many of the groups here, as they are the largest scale of the hierarchy that may be gravitationally bound (although they are not necessarily so, e.g. Bastian \& Gieles~2006), thus potentially long lived.  So, any scale imprinted on them during their formation may remain, thus explaining why old globular clusters (i.e. Jord{\'a}n et al.~2005) and young massive clusters (e.g. Bastian et al.~2005, Scheepmaker et al.~2007) have a similar characteristic size.  Put another way, if this scale separates from the hierarchy (i.e. structures on this scale remain bound while structures on larger scales diffuse, and groups on smaller scales disrupt or merge) then the resulting clusters will bear the imprint of being a scale within the hierarchy, namely possess a log-normal size distribution.


\section{Summary}\label{sec:conclusions}

We have studied star-forming regions in M33 using the photometric catalogue of Massey et al.~(2006), and applying an objective finding and categorising algorithm, fractured Minimum Spanning Trees (fMST).  Using colour and magnitude cuts we have isolated young massive stars, and constructed a minimum spanning tree for these sources.  We then fractured the tree using thirteen breaking distances, $19 \le$~\db~$\le 252$~pc, to isolate star-forming regions of various sizes.  

A study of their properties allowed us to draw the following conclusions:

\begin{enumerate}
\item The radius distribution of the groups does not have a power-law distribution, but instead is well described by a log-normal function.   The characteristic size (i.e. mean value) of the radius distribution is determined by the \db~ and \nmin~ adopted.
\item Contrary to previous claims, we do not find any characteristic scale in the star-forming groups sizes.  Previous claims of such a distinctive size scale are shown to be the result of a combination of resolution and selection effects.
\item The luminosity function of the regions is well described by a power-law of the form $NdL \propto L^{-\alpha}$ where $\alpha = 2.00\pm0.03$.  This is the same as found for star cluster populations and similar to the mass function of giant molecular cloulds.
\item There is a strong spatial correspondence between bright \hii regions and the stellar groups.  This suggests that these \hii regions are powered by multiple high-mass stars.  Faint \hii regions, however, are often not found to be associated with the stellar groups found here, suggesting that they are powered by single massive stars or multiple lower mass stars that do not pass our selection criteria.
\item We find a distinct truncation of the galactocentric distribution of the regions at 4~kpc, the same as was found for GMCs in this galaxy (Rosolowsky et al.~2007).   However, molecular gas and H{\sc ii} regions are both found outside this radius, suggesting a change in the structure of the star-forming ISM which is not seen in the radial profiles of the molecular gas or star-formation tracers.
\item The spatial distributions of the GMCs, H{\sc ii} regions, and star-forming groups are found to be highly correlated up to scales of $>$ 100~pc.
\item The size distribution of the regions is consistent with previous suggestions that the distribution of star-formation in galaxies is hierarchical and fractal.
\item The method discussed here is comparable to that of the Gaussian smoothing method of Elmegreen \& Elmegreen~(2001) and Elmegreen et al.~(2006), where our parameter \db~ plays a similar role to their smoothing length.  However, our method retains the information of the individual sources within each region allowing for further detailed study.
\item By the construction of simple three dimensional fractals, projected onto a two dimensional plane, we have shown that if a particular size scale, i.e. \db, is chosen for study, the resulting size distribution of the groups will naturally be a log-normal distribution.  If, however, one studies the number of groups found for each \db, the resulting distribution will be a power-law as predicted by previous studies.  Both effects are seen in the star-forming groups of M33, arguing for a hierarchical distribution of the star-forming regions.  Finally, we suggest that this affect may explain the observed log-normal size distribution of star clusters.

\end{enumerate}

We have limited our analysis  to very young (dominated by stars $< 20$~Myr old) groups.  However, it has been noted that these young groups are often contained within larger and older groups - star complexes (e.g. Efremov 1979, 1995).  These groups can be found using older stellar population tracers, such as Cepheid variables.  Such older stars are mostly not found in compact groups, presumably due to the gravitationally unbound nature of star-forming groups (which is considered in detail in Gieles, Bastian, \& Ercolano 2007).  However, these older stars are often associated over scales of a few hundred parsecs (Efremov 1995) and will be studied in detail for M33 in a future work.

We have seen that star-formation in M33 appears to be hierarchical, with structures present on a multitude of scales, from clusters and OB-assocations, to stellar complexes with sizes of hundreds of parsecs.  The question then is how far down in scale does this hierarchy proceed?  Higher-resolution studies are thus required.  These studies must be shifted out of the optical, as the dynamical timescale of these systems decreases as the scale decreases, and therefore we must trace ever-younger populations which are generally more effected by extinction.  Near and mid-IR observations of active star-forming sites in the galaxy offer a unique chance to study the distribution of star-forming sites at parsec and sub-pc scales.

\section*{Acknowledgments}

We thank the referee, Phil Massey, and Deidre Hunter for their comments and suggestions which improved the paper.  Bruce Elmegreen is thanked for insightful discussions on the size distribution of star-forming regions. NB gratefully acknowledges the hospitality of the Harvard-Smithsonian Center for Astrophysics, where a significant part of this work took place, and is supported by a PPARC post-doctoral fellowship.  BE is partially supported by {\it Chandra} grants GO6-7008X and GO6-7009X.  ER's work is supported by NSF grant AST-0502605.  YuE is grateful for support to the RFBR (project 06-02-16077).

\bsp
\label{lastpage}
\end{document}